\documentclass[%
 preprint,
 amsmath,amssymb,
]{revtex4-1}

\usepackage{graphicx}
\usepackage{amssymb}
\usepackage{setspace}
\usepackage[utf8]{inputenc}
\usepackage{array}
\usepackage{natbib}
\usepackage{amsmath}
\usepackage{multirow}
\usepackage{hyperref}
\usepackage{rotating}
\usepackage{subcaption}
\usepackage{adjustbox}
\usepackage{threeparttable}
\usepackage{booktabs}

\usepackage{siunitx}
\usepackage{color}
\definecolor{redcolor}{rgb}{1.0,0.,0.}

\definecolor{bluecolor}{rgb}{0,0.,1}

\begin{document}

\preprint{}

\title{Leveraging GANs for citation intent classification and its impact on citation network analysis}

\author{Davi A. Bezerra$^1$,  Filipi N. Silva$^2$ and Diego R. Amancio$^1$}

\affiliation{
$^1$Institute of Mathematics and Computer Science, University of S\~ao Paulo, S\~ao Carlos, Brazil\\\\
$^2$Observatory on Social Media, Indiana University, Bloomington, IN, USA \\
}

\newpage


\begin{abstract}
Citations play a fundamental role in the scientific ecosystem, serving as a foundation for tracking the flow of knowledge, acknowledging prior work, and assessing scholarly influence. In scientometrics, they are also central to the construction of quantitative indicators. Not all citations, however, serve the same function: some provide background, others introduce methods, or compare results. Therefore, understanding citation intent allows for a more nuanced interpretation of scientific impact. In this paper, we adopted a GAN-based method to classify citation intents. {Our results revealed that the proposed method achieves competitive classification performance, closely matching state-of-the-art results with substantially fewer parameters. This demonstrates the effectiveness and efficiency of leveraging GAN architectures combined with contextual embeddings in intent classification task}. We also investigated whether filtering citation intents affects the centrality of papers in citation networks. Analyzing the network constructed from the unArXiv dataset, we found that paper rankings can be significantly influenced by citation intent. All four centrality metrics examined -- degree, PageRank, closeness, and betweenness -- were sensitive to the filtering of citation types. The betweenness centrality displayed the greatest sensitivity, showing substantial changes in ranking when specific citation intents were removed.

\end{abstract}

\maketitle








\frenchspacing



\doublespacing

\section{Introduction}
\label{Introdução}

\noindent Citations play a fundamental role in scientific production, 
connecting research works to the existing body of academic literature 
and situating new advancements within the context of previous discoveries. 
Citation analysis has been widely employed to detect emerging research topics, 
measure the impact of publications, authors, institutions, and improve 
indicators such as the impact factor, in addition to powering more 
informative indexing systems and automatic summarization tools 
\cite{teufel2006annotation, jebari2021use, liho2008use}. 
However, not all citations carry the same weight or serve equivalent functions. 
The role of a citation can vary depending on the context in which it is employed, 
such as justifying the relevance of a research question, describing a method, 
comparing results, or even refuting previous work. This specific purpose, 
often referred to as the \textit{function} \cite{jurgens2018} or 
\textit{intent} \cite{cohan2019structural} of a citation, 
is central to a richer and more qualitative analysis of academic impact.

Having an intent classification is critical for improving the automated analysis of academic literature and the measurement of scientific impact \cite{small2018}. Other applications of citation intent include improved search experience \cite{michael1975resultson}, information retrieval \cite{ritchie2008}, summarization \cite{cohan2015matching}, and studying the evolution of scientific fields \cite{jurgens2018}. Earlier works were focused on creating more fine-grained categories, going as far as defining schemes with 35 \cite{garzone1997automated} and 12 \cite{teufel2006annotation} categories for scientific arguments. More recent works, however, have focused on creating more concise categories. For example, ACL-ARC \cite{jurgens2018} proposes a 6-class intent categorization scheme: Background, Motivation, Uses, Extension, Comparison or Contrast, and Future. SciCite \cite{cohan2019structural} is even more restrictive, dropping or combining small fine-grained classes to provide a more concise 3-class annotation scheme: Background, Method, and Result.

Before the advent of machine learning models, citation intention classification was already done manually, focusing on just one article to understand how citations occurred \cite{swales1986citation, white2004citation}. With the advancement of automatic techniques for intention classification, annotated databases, and computational power, it became possible to automatically identify the intention of authors when citing other authors in an automatic way \cite{teufel2006automatic, valenzuela2015identifying}.

{Given the scarcity of large-scale annotated citation intent datasets, semi-supervised learning techniques are particularly well-suited for this task as they can leverage abundant unlabeled data alongside limited labeled samples. In this study, we adopted a semi-supervised GAN framework, specifically GAN-BERT enhanced with SciBERT embeddings (SS-cGAN + SciBERT), to effectively address this challenge. This approach exploits the generative adversarial network’s ability to improve generalization by integrating unlabeled examples during training, thereby overcoming the limitations of purely supervised methods. Our model achieved competitive performance on benchmark datasets, reaching an F1 score of 88.74 on SciCite and 81.75 on ACL, outperforming several baselines and demonstrating robust generalization despite limited labeled data. Although performance on the more challenging 3C dataset was lower, the overall results highlight the potential of semi-supervised GANs to enhance citation intent classification in realistic, data-constrained scenarios.}

A second contribution of this work is the analysis of how citation intent affects citation network analysis.
Most existing studies treat citation intent classification and citation network analysis as separate tasks, either focusing solely on intent classification or analyzing citation networks without considering the underlying intent of each citation. Recent studies \cite{ghosh_2020, gupta2024sentiment} have advanced this area by incorporating sentiment-aware PageRank, which adjusts the weight of each citation according to its polarity—assigning greater influence to positive citations and reducing the impact of critical ones—to better reflect how different citation sentiments affect the ranking of papers and authors. However, these approaches do not leverage citation intent, which provides a more comprehensive categorization of citations beyond mere sentiment. To address this gap, our research proposes an integrated methodology that combines citation intent classification with citation network analysis. We apply network centrality measures that distinguish citations based on their academic function, allowing for a nuanced evaluation of how filtering specific citation intents influences rankings. Our results revealed that filtering citations by intention can change the interpretation of which papers are the most central in a paper citation network.

%



\section{Related Works}
\label{TheoreticalFramework}

The classification of citation intent has evolved significantly 
from manual analyses to sophisticated machine learning approaches. 
Initial works such as Garfield et al. \cite{garfield1965can} laid the 
groundwork by identifying primary citation motivations, 
including alerting researchers to relevant works, acknowledging foundational contributions, 
and pointing readers to background literature. 
Weinstock \cite{weinstock_1971} expanded this categorization, 
providing detailed insights into various citation purposes. 
Likewise, Moravcsik and Murugesan \cite{moravcsik_1975} underscored 
the importance of classifying citations to better understand scholarly communication.

With the advent of computational methods, citation intent classification 
transitioned from manual efforts to automated systems. 
Early computational models were predominantly rule-based \cite{based_rules_2000} 
or employed straightforward statistical methods such as multinomial 
naïve Bayes and Support Vector Machines (SVM) \cite{svm_naive_bayes_2010}.
Ensemble methods and nearest-neighbor algorithms also featured prominently 
in early computational approaches \cite{dong-schafer-2011-ensemble, teufel2006automatic}.

A major step forward came from the work of Jurgens et al. \cite{jurgens2018}, 
who proposed a robust annotation schema classifying citations into six distinct 
categories: background, motivation, uses, extension, comparison or contrast, 
and future works. Leveraging lexical, topical, structural, and syntactic features, they achieved substantial improvements over previous models, 
highlighting how different citation frames could predict the citation impact and reveal disciplinary trends.

Cohan et al. \cite{cohan2019structural} further advanced the field by 
introducing multitask learning into citation intent classification. 
Their model employed structural scaffolds—tasks such as citation worthiness 
and section title prediction to enhance classification performance significantly. This approach outperformed previous methods by effectively capturing the inherent structural properties of scientific discourse, notably on the ACL-ARC dataset \cite{jurgens2018} and the newly introduced SciCite dataset.

The emergence of transformer-based models marked another leap forward. 
Beltagy et al. \cite{beltagy_etal_2019_scibert} introduced SciBERT, 
a variant of BERT specifically trained on scientific literature. 
SciBERT demonstrated superior performance compared to general-domain models 
on citation intent tasks, highlighting the importance of domain-specific pretraining. 
Mercier et al. \cite{mercier2020impactcite} built upon this trend, 
introducing \emph{ImpactCite}, an XLNet-based model that captured deeper 
contextual and structural relationships, significantly 
outperforming simpler architectures such as CNNs and RNNs.

More recent innovations continued refining these transformer-based methodologies. 
VarMAE \cite{hu2022varmae} employed a Variational Masked Autoencoder and 
Context Uncertainty Learning (CUL) to enhance domain adaptation. 
This approach provided strong results in low-resource settings and 
effectively handled domain-specific contexts on ACL-ARC and SciCite datasets.

Lahiri et al.~\cite{lahiri2023citeprompt} proposed CitePrompt, 
introducing a prompt-based learning strategy that reframed 
citation intent classification into a masked language modeling task. 
CitePrompt used prompt templates and verbalizers, 
reducing reliance on extensive labeled datasets and external resources. 
Its effectiveness in few-shot and zero-shot learning scenarios demonstrated remarkable 
adaptability and robustness, further advancing the capabilities of citation intent classifiers.

Parallel to advancements in citation intent classification, 
research on citation networks has emphasized the utility of 
centrality measures for understanding scientific communication. 
Betweenness centrality, as explored by Leydesdorff \cite{betweenness2007loet}, 
highlights journals that bridge different scientific disciplines, 
acting as connectors within citation networks. 
Similarly, Yan et al. \cite{centrality2009yan} demonstrated 
the relevance of centrality measures such as degree, closeness, 
betweenness, and PageRank in assessing author influence within 
coauthorship networks. Their findings reveal significant 
correlations between centrality measures and traditional 
bibliometric indicators such as citation counts, suggesting that 
centrality metrics effectively capture both direct and indirect scholarly impact.

Overall, the evolution from manual analyses to advanced deep learning models 
underscores ongoing advancements in understanding citation intent. 
Each methodological leap, from structured annotation schemas and multitask 
frameworks to transformer architectures and prompt-based learning, 
illustrates the growing sophistication of citation intent classification tools, 
contributing to scholarly communication analysis. 
Despite these advancements, the integration of citation intent classification
with citation network analysis remains largely unexplored. 
Understanding how citation intent influences network metrics could enhance the 
interpretation of scientometric indicators and provide new insights 
into the dynamics of scientific knowledge dissemination.

\section{Materials and Methods}
\label{Materials}

This section provides a detailed overview of the methodology developed and refined throughout this study. The main objective is to investigate citation intent classification and its significance in enhancing the understanding of each citation's relevance within a citation network. To achieve this,
we adopt a semi-supervised deep learning model to automatically classify citation intentions. In a second step, we re-evaluate the relevance of papers by filtering citations based on their intent, under the assumption that more purposeful citations -- rather than merely contextual ones -- provide a better assessment of a paper’s significance.

%

To assess the performance of the proposed classification model and analyze citation impact across diverse scientific domains, we employed a combination of benchmark datasets and large-scale scholarly publication corpora. For the citation intent classification task, we used three well-established datasets: ACL-ARC, SciCite, and the 3C Shared Task Dataset. Additionally, the Unarxiv dataset was leveraged to construct the citation network and facilitate model inference, enabling the analysis of the impact of filtering specific citation intents.

\subsection{Citation Intent Classification Datasets}

\begin{enumerate}
    \item \emph{ACL-ARC}: This dataset is a collection of citation intents based on a subset of papers from the ACL Anthology Reference Corpus \cite{bird_etal_2008_acl}. The dataset comprises 1,941 annotated citation instances drawn from 186 papers in the computational linguistics and natural language processing domains, grouped into six distinct intent classes (see Table \ref{tab:citation_functions}). Each citation instance contains the citation context and details about both the citing and cited papers. The intent categories exhibit an imbalance, with the majority being background, uses, and compares or contrasts, while motivation, continuation, and future are underrepresented as can be seen in Table \ref{tab:datasets}.
    
    \item \emph{Scicite}: In 2019, \cite{cohan2019structural} introduced the SciCite dataset, which addresses limitations in existing citation intent datasets by providing a larger, more general-domain collection of annotated citations. It simplifies citation intent classification into three categories: background, method, and result. Table \ref{tab:citation_intent} provides detailed descriptions of each category. This new scheme merges several categories from previous datasets, such as \cite{tacl_a_2018_acl}, into a single background category, arguing that this better reflects the structure of scientific discourse across diverse domains. The dataset includes two additional tasks: Section Title Prediction with $91,412$ instances and r labels, and Citation Worthiness Prediction with 73,484 instances labeled and 
    2 labels. The task involves training a model to flag sentences that require a citation by treating those containing citation markers (e.g., “[12]” or “Lee et al. (2010)”) as positive examples and sentences without markers as negative, thereby capturing the distinctive linguistic patterns of citation-worthy text \cite{cohan2019structural}.
    
    \item \emph{3C Shared Task Dataset~\footnote{https://www.kaggle.com/c/3c-shared-task-purpose-v2}}: The 3C Challenge dataset is a subset of the ACT dataset \cite{knoth_act}. 
    It supports the classification of citation contexts into six categories: background, uses, compares and contrasts, motivation, extension, and future. The dataset consists of 3,000 training instances and 1,000 testing instances, though the test category data is not publicly available.
\end{enumerate}

\begin{table}[ht]
\centering
\caption{Overview of citation datasets. The categories represent possible citation classifications. \textit{D (\%)} indicates the percentage of citations falling into each category, and \textit{NI} denotes the total number of instances in each category.}
\label{tab:datasets}
\begin{tabular}{llcc}
    \hline
\textbf{Dataset} & \textbf{Categories} & \textbf{\textit{D} (\%)} & \textbf{\textit{NI}} \\
    \hline
\multirow{6}{*}{ACL-ARC}
 & Background        & 51 & \num{997} \\
 & Compare/contrast  & 18 & \num{351} \\
 & Extends           &  4 & \num{72}  \\
 & Future work       &  4 & \num{69}  \\
 & Motivation        &  5 & \num{88}  \\
 & Uses              & 19 & \num{364} \\
    \hline
\multirow{6}{*}{3C Shared Task \ }
 & Background        & 55 & \num{1648} \\
 & Compare/contrast  & 12 & \num{368}  \\
 & Extension         &  6 & \num{171}  \\
 & Future            &  2 & \num{62}   \\
 & Motivation        &  9 & \num{276}  \\
 & Uses              & 16 & \num{475}  \\
    \hline
\multirow{3}{*}{SciCite}
 & Background        & 58 & \num{6375} \\
 & Method            & 29 & \num{3154} \\
 & Result comparison & 14 & \num{1491} \\
    \hline
\end{tabular}
\end{table}

All three datasets  vary significantly in their properties, reflecting distinct focuses as can be seen in Table \ref{tab:dataset-atributes}. ACL-ARC and 3C Shared Task categorize citation intents into six nuanced classes, whereas SciCite simplifies classification into three broader categories, emphasizing general applicability. The average context lengths are comparable (33-36 words), but SciCite stands out with a maximum length of 510 words, offering richer contextual information compared to ACL-ARC (178 words) and 3C Shared Task (117 words). Dataset sizes also differ, with SciCite being the largest (8,243 training and 1,851 test samples), providing robust resources for model training. In contrast, ACL-ARC (1,688 training and 139 test samples) and 3C Shared Task (2,000 training and 500 test samples) focus on domain-specific classification. These differences make the datasets complementary, with SciCite suited for general applications and the others for more detailed analyses.

\begin{table}[htb]
    \centering
    \caption{Overview of datasets with their sentence properties.}
    \footnotesize  
    \begin{tabular}{l c c c c c c}
    \hline
    \textbf{Dataset}      & \textbf{Classes}   & \textbf{Avg lengths} & \textbf{Max lengths}  & \textbf{Train samples} & \textbf{Test samples} \\ \hline
    ACL-ARC                 & 6       & 33   & 178       & 1,688     & 139   \\ 
    SciCite                 & 3       & 35   & 510       & 8,243     & 1,851    \\ 
    3C Shared Task          & 6       & 36   & 117       & 3,000     & 1000  \\ \hline
    \end{tabular}
    \label{tab:dataset-atributes}
\end{table}

\begin{table}[ht]
    \centering
    \caption{Definitions and representative examples of the six citation intent functions a citation may serve with respect to a cited paper $P$) in the ACL-ARC dataset (adapted from ref.~\cite{jurgens2018}).}
    \label{tab:citation_functions}
    \resizebox{\textwidth}{!}{%
    \begin{tabular}{l p{0.35\textwidth} p{0.45\textwidth}}
        \hline
        \textbf{Class} & \textbf{Description} & \textbf{Example} \\ 
        \hline
        BACKGROUND
            & Provides relevant information for this domain.
            & “This is often referred to as incorporating deterministic closure” (Dörre, 1993). \\
        \hline
        MOTIVATION
            & Illustrates the need for data, goals, methods, etc.
            & “As shown in Meurers (1994), this is a well-motivated convention \ldots” \\
        \hline
        USES
            & Uses data, methods, etc., from \(P\).
            & “The head words can be automatically extracted in the manner described by Magerman (1994).” \\
        \hline
        EXTENSION
            & Extends \(P\)’s data, methods, etc.
            & “\ldots we improve a two-dimensional multimodal version of LDA” (Andrews et al., 2009). \\
        \hline
        COMPARISON/CONTRAST
            & Expresses similarities or differences to \(P\).
            & “Other approaches use less deep linguistic resources (e.g., POS-tags)” (Stymne, 2008). \\
        \hline
        FUTURE
            & Indicates a potential avenue for future work on \(P\).
            & “\ldots but we plan to do so in the near future using the algorithm of Littlestone and Warmuth (1992).” \\
        \hline
    \end{tabular}%
    }
\end{table}

\begin{table}[ht]
    \centering
    \caption{Definitions and representative examples of citation intent categories in the SciCite dataset, based on \cite{cohan2019structural}.}
    \label{tab:citation_intent}
    \resizebox{\textwidth}{!}{%
    \begin{tabular}{l p{0.35\textwidth} p{0.45\textwidth}}
        \hline
        \textbf{Class} & \textbf{Description} & \textbf{Example} \\
        \hline
        BACKGROUND
            & The citation provides background context, elaborating on a problem, concept, approach, topic, or the significance of the problem in the field.
            & Recent evidence suggests that co‐occurring alexithymia may explain deficits [12]. \newline
              Locally high‐temperature melting regions can act as permanent termination sites [6–9]. \newline
              One line of work is focused on changing the objective function (Mao et al., 2016). \\
        \hline
        METHOD
            & The citation references a method, tool, approach, or dataset used in the study.
            & Fold differences were calculated using a mathematical model described in [4]. \newline
              We use Orthogonal Initialization (Saxe et al., 2014). \\
        \hline
        RESULT
            & The citation compares the paper’s results/findings with those of prior research.
            & Weighted measurements were superior to T2‐weighted contrast imaging, aligning with previous studies [25–27]. \newline
              Similar results to our study were reported by Lee et al. (2010). \\
        \hline
    \end{tabular}%
    }
\end{table}

\subsection{Large-Scale Scholarly Publication Dataset}

\begin{figure}[htb]
\centering
\includegraphics[width=1.0\linewidth]{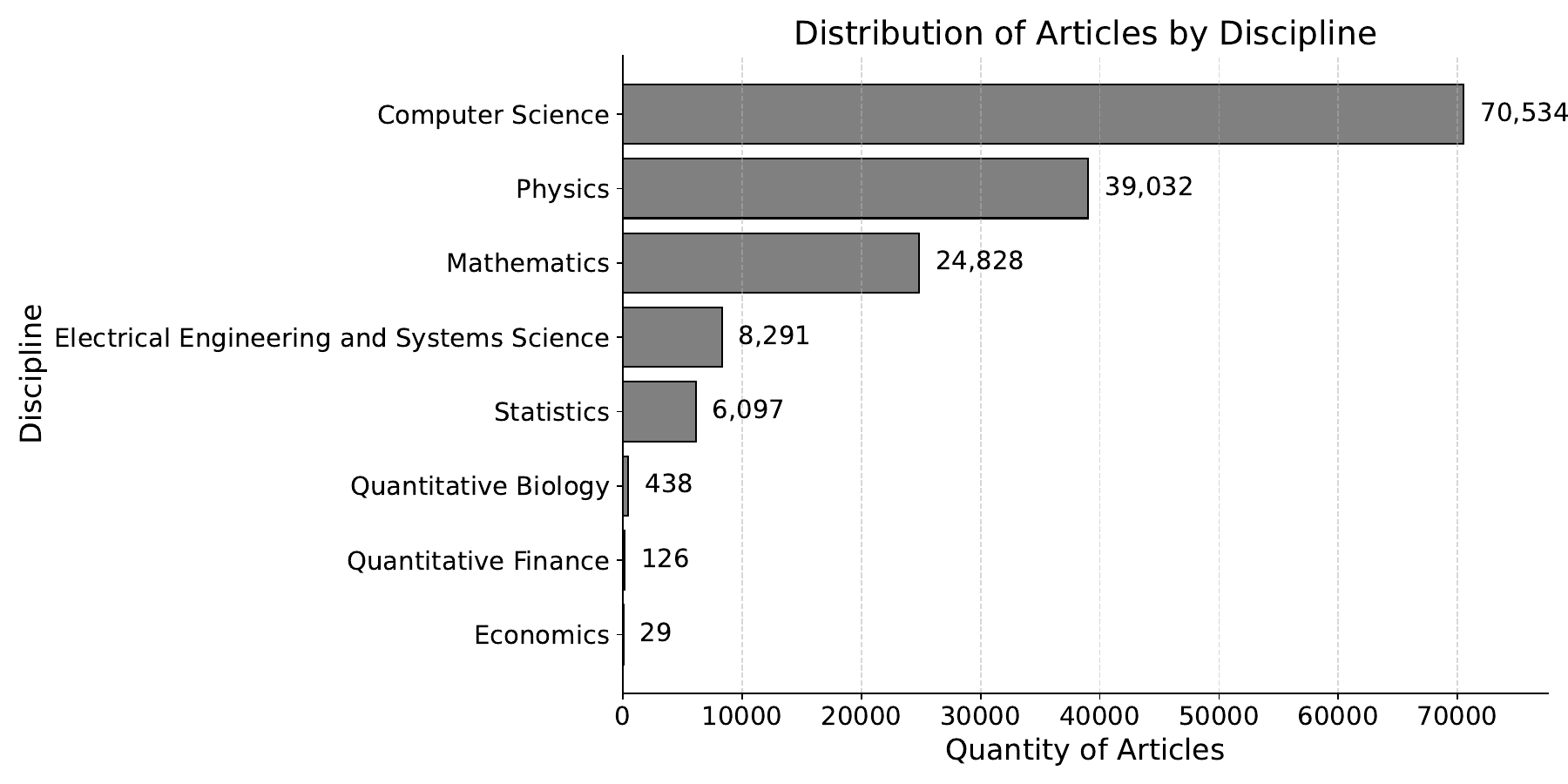}
\caption{\label{fig:distribution_articles }Distribution of articles across disciplines in the unarXiv dataset. The dataset demonstrates a significant imbalance, with computer science, physics, and mathematics accounting for 47\%, 26\%, and 17\% of the total articles, respectively.}
\end{figure}

To construct the citation network and analyze the impact of different citation intentions on its structure, we used the unarXiv dataset \footnote{https://github.com/IllDepence/unarXive}~\cite{saier-2020-unarXive}. UnarXiv serves as a robust resource for bibliometric analysis and Natural Language Processing (NLP) tasks, encompassing a comprehensive collection of academic publications spanning 32 years across diverse disciplines, including physics, mathematics, computer science, and related fields. The dataset contains over 1.8 million articles, each with detailed document structures, enabling in-depth exploration of citation networks, scientific trends, and the evolution of research topics over time. However, it is important to note that the dataset exhibits a significant imbalance in disciplinary coverage. While computer science, physics, and mathematics collectively account for over 90\% of the total articles, other disciplines are underrepresented due to their limited presence on arXiv, as illustrated in Figure \ref{fig:distribution_articles }. Specifically, computer science constitutes 47\% of the dataset, followed by physics at 26\% and mathematics at 17\%. This disparity highlights the dataset's bias toward disciplines that predominantly utilize arXiv as a preprint repository.

Prior to conducting the analysis, two critical preprocessing steps were undertaken: data preparation and inference execution. The unarXiv dataset is provided in JSONL (JSON Lines) format, which required the extraction of essential metadata and content for subsequent analysis. Key information, such as citation contexts, article titles, the specific sections in which citations occurred, and additional metadata relevant to the study, were systematically extracted. Following this, the data were transformed into a CSV (Comma-Separated Values) format to streamline data manipulation and ensure compatibility with the inference pipeline.

\subsection{Methodology for Citation Intent Classification}

{In this study, we adopted GAN-BERT for the classification task. Motivated by the fact that understanding citation intent adds semantic depth to citation links -- revealing whether a reference serves as foundational background, a source of methods or data, a target of critique, or supporting evidence -- GAN-BERT is particularly suited for scenarios with limited annotated data, effectively leveraging semi-supervised learning techniques to integrate both labeled and unlabeled data during model training~\cite{gan_cite_2025, silva_etal_2023, auti_etal_2022}. We apply GAN-BERT to scientific databases with few annotated examples, a common situation in citation analysis tasks, since the scarcity of labeled data hampers efficient learning and restricts the generalization of traditional supervised classification models.}

GAN-BERT \cite{croce-etal-2020-gan} is an architecture that combines BERT \cite{devlin2019bert} with semi-supervised learning techniques, specifically leveraging the strengths of SS-GAN. By integrating the contextual representation capabilities of BERT with the adversarial training framework of GAN, GAN-BERT enhances performance in sentence classification tasks. This hybrid model is particularly advantageous in scenarios where labeled data is limited, as it effectively utilizes both supervised and unsupervised learning paradigms.

BERT captures the intricate contextual relationships between words in a text using a multi-layer bidirectional attention mechanism. This mechanism allows BERT to generate highly contextualized embeddings, providing meaningful representations for individual tokens and comprehensive representations at the sentence level. In the specific context of citation intent classification, BERT processes the input citation context by tokenizing it into a sequence \( s = (h_{\text{CLS}}, h_{s1}, \ldots, h_{sn}, h_{\text{SEP}}) \), where \( h_{\text{CLS}} \) represents the special token summarizing the entire citation context, \( h_{s1}, \ldots, h_{sn} \) are the embeddings for individual tokens, and \( h_{\text{SEP}} \) is a special separator token marking the end of the input sequence.

In this study, we employ SciBERT \cite{beltagy_etal_2019_scibert}, a domain-specific variant of BERT, to generate vector representations for citation intent classification. SciBERT is uniquely designed for processing scientific text and was pre-trained on a large corpus of 1.14 million academic papers spanning the computer science and biomedical domains, amounting to 3.1 billion tokens. Its academic focus and specialized vocabulary make SciBERT exceptionally well-suited for analyzing scholarly content. Among its available configurations, the \textit{scivocab-uncased} version was selected for our experiments.  While this study employs SciBERT \cite{beltagy_etal_2019_scibert}, the proposed methodology is not limited to this model and can be extended to other transformer-based architectures such as XLNet or Modern BERT.

\begin{figure}[htb]
    \centering
    \includegraphics[width=1.0\linewidth]{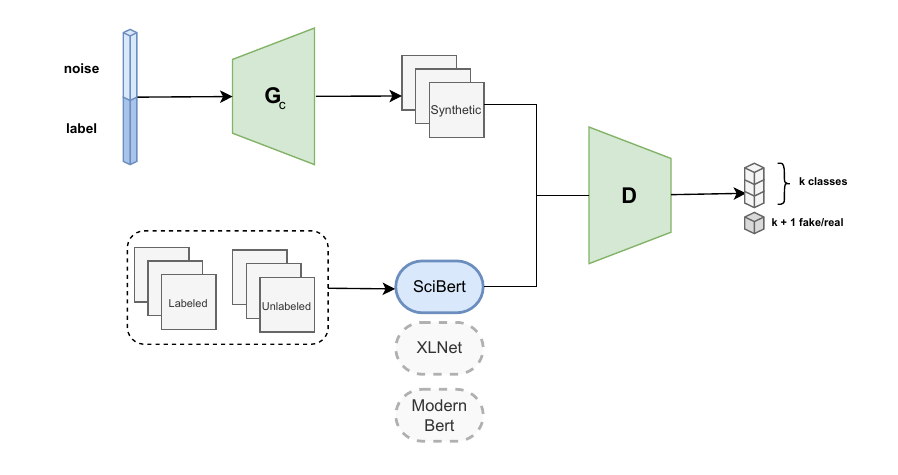}
    \caption{Architecture of cGAN-SciBERT. The model integrates SciBERT with an SS-GAN framework, comprising a generator (\( G_c \)) and a discriminator (\( D \)). The generator produces synthetic examples from noise and a class-specific vector, while the discriminator classifies real examples and detects fake ones.}
            \label{fig:gan-bert-architecture}
\end{figure}

The structure of the proposed conditional GAN-BERT, or cGAN-SciBERT, is illustrated in Figure~\ref{fig:gan-bert-architecture}, showcasing how SciBERT is combined with a semi-supervised GAN framework consisting of a conditional generator ($G_c$) and a discriminator ($D$). As shown in the Figure~\ref{fig:gan-bert-architecture}, it is possible to use other architectures to generate the representations that will be used by the discriminator, suggesting flexibility and extensibility of the proposed framework to incorporate alternative transformer-based models or embedding techniques.

In the original GAN-BERT framework, the generator primarily created adversarial examples to enrich the discriminator’s training process, providing limited insights into its role in representation learning. In our proposed approach, we introduce a conditional generator specifically designed to produce representations aligning closely with the characteristics of particular classes or dataset attributes, thus offering deeper insights into how these representations are formed.

The conditional generator $G_c$, implemented as a Multi-Layer Perceptron (MLP), generates synthetic examples $h_{\text{fake}}$ from a noise vector $z$, sampled from a Gaussian distribution, combined with a class-specific conditional vector. These synthetic examples are constructed to closely resemble the real data distributions for specific categories, challenging the discriminator to distinguish between authentic and generated samples. By producing high-quality synthetic data, the generator indirectly supports the discriminator in capturing nuanced data features, particularly beneficial when labeled examples are scarce. 

The discriminator $D$, also implemented as an MLP, serves a dual purpose. It classifies real examples $h_{\text{CLS}}$, extracted from the [CLS] token of the SciBERT encoder, into one of the predefined $k$ categories. Simultaneously, it identifies synthetic examples produced by the generator as belonging to a separate, distinct class $(k+1)$. Through this dual functionality, the discriminator leverages both labeled and unlabeled data structures effectively, enhancing its overall learning capability.

The training of GAN-BERT follows an adversarial process with competing objectives for the generator and discriminator. The generator seeks to create synthetic examples indistinguishable from real data, thus minimizing detection as fake by the discriminator. Conversely, the discriminator aims to maximize its accuracy in distinguishing real from synthetic examples and correctly classifying genuine data into their respective categories. The labeled examples contribute to the supervised loss $L_{\text{sup}}$, while unlabeled data and generated examples influence the unsupervised loss $L_{\text{unsup}}$. This adversarial dynamic refines both the discriminator’s classification skills and the underlying SciBERT representations through backpropagation.

Upon completing training, the generator is discarded, retaining only the fine-tuned SciBERT model and discriminator for inference. This ensures enhanced classification performance without incurring additional computational costs during deployment. The code, data, and trained model weights are available at this \href{https://github.com/davialvb/intent-citation-network}{link}.

\subsection{Effects of filtering citations by intention in citation networks}

The second objective of this paper is to analyze the impact of selecting specific citation intentions on the structure of citation networks. This is important because not all citation intentions are equally relevant, and their significance may vary depending on the analytical purpose. For instance, if one aims to assess the importance of a paper based on its use as a methodological reference, it would be more meaningful to consider only citations that reflect this specific intent. This approach is consistent with other citation filtering methods, such as those that assess author relevance by considering their specific contributions to papers~\cite{brito2023analyzing,correa2017patterns}.

Paper relevance in citation networks is evaluated based on centrality measures~\cite{bagrow2024working}. While centrality metrics have been used in many contexts where network can model real-world systems, including semantic and social networks~\cite{correa2019word,brito2020complex}, here we use centrality metrics to quantify the importance of papers in a subfield.
We employed the following metrics that are commonly used in network analysis: degree, betweenness, closeness and PageRank~\cite{bagrow2024working}. A more detailed explanation of these metrics are available in the Supplementary Information. 
All metrics are computed using adjacency matrices derived from the citation data. For each node (paper), the centrality values are calculated to assess its structural position and potential influence within the network. By comparing these metrics across different citation intents, the study identifies whether certain types of citations (e.g., background, method, or result) are associated with higher centrality values, suggesting a more significant role in shaping scientific discourse.

To construct a meaningful citation network from the unarXiv dataset, a citation filtering process is applied to ensure that only relevant and high-impact citations are included. The network analysis consists of three main steps: 

\begin{enumerate}

\item \emph{Intent-Based Filtering}: citations are categorized using a citation intent classifier trained with the SciCite dataset, employing GAN-BERT enhanced by SciBERT embeddings. The SciCite dataset was selected due to its coarse-grained, general-domain annotation scheme, which effectively captures the principal citation intents into three clear and meaningful categories: \emph{BACKGROUND}, providing context or motivation; \emph{METHOD}, indicating direct usage of methods or tools; and \emph{RESULT}, emphasizing comparison or analysis of outcomes. Utilizing SciCite enables clearer insights into citation dynamics and allows for effective visualization of the network-level impacts resulting from intent-based citation filtering.

\item \emph{Implementation and Analysis}: the citation network is represented as a directed graph \( G = (V, E) \), where nodes \( V \) correspond to papers, and edges \( E \) represent citation links. The adjacency matrix \( A \) is used to compute centrality measures, with element \( a_{ij} = 1 \) if paper \( i \) cites paper \( j \), and 0 otherwise. 

\item \emph{Visualizing Citation Networks}: network visualizations are generated using Helios-Web~\footnote{https://github.com/filipinascimento/helios-web}, illustrating the structural roles of key papers within their citation contexts~\cite{silva2016citationnetwork}.

\end{enumerate}

\section{Results}
\label{Results}

This section presents a comprehensive evaluation of the models employed in this study, with particular emphasis on the comparative performance of our proposed approach, cGAN-SciBERT, against established baseline methods. We begin by analyzing classification metrics to assess predictive accuracy and examine the distribution of mean classification errors to understand the model's reliability when applied to large-scale scientific corpora. We also investigate the structure of the underlying citation network through the lens of centrality measures, with a specific focus on how filtering citations based on their rhetorical intent impacts the topological and semantic properties of the network.

\subsection{Model Performance}
\label{sec:model_performance}

Table \ref{tab:f1_results} summarizes the results of our experiments, comparing the performance of the proposed SS-cGAN + SciBERT method against several baseline models on the SciCite, ACL, and 3C datasets. Performance is evaluated using the F1 score, a key metric for assessing the balance between precision and recall in citation intent classification tasks. Details on the hyperparameter settings used during model training are provided in the Supplementary Information section.

\begin{table}[htb]
\centering
\begin{adjustbox}{max width=\textwidth}
\begin{tabular}{l c c c}
\hline
\textbf{Model}                       & \textbf{SciCite} & \textbf{ACL}   & \textbf{3C (Public F1 / Private F1)} \\ \hline
ImpactCite                           & \textbf{88.93}   & -              & - \\ \hline 
\textbf{SS-cGAN + SciBERT}           & 88.74            & \textbf{81.75} & 26.22 / 23.21  \\ \hline
\textbf{SS-cGAN + XLNet}             & 87.61            & 75.60          & 23.95 / 23.08 \\ \hline
CitePrompt                           & 86.33            & 68.39          & - \\ \hline
VarMAE                               & 86.32            & 76.50          & - \\ \hline
SciBERT Finetune                     & 85.49            & 70.98          & - \\ \hline
BiLSTM-Attn w/ ELMo + both scaffolds & 84.0             & 67.9           & 30.3 / 26.1 \\ \hline
3C Shared Task Best Submission       & -                & -              & \textbf{33.9 / 27} \\ \hline
\end{tabular}
\end{adjustbox}
\caption{Performance of our proposed method against different baseline methods for citation intent classification across multiple datasets.}
\label{tab:f1_results}
\end{table}

\begin{figure}[htb]
    \centering
    \includegraphics[width=130mm,scale=0.5]{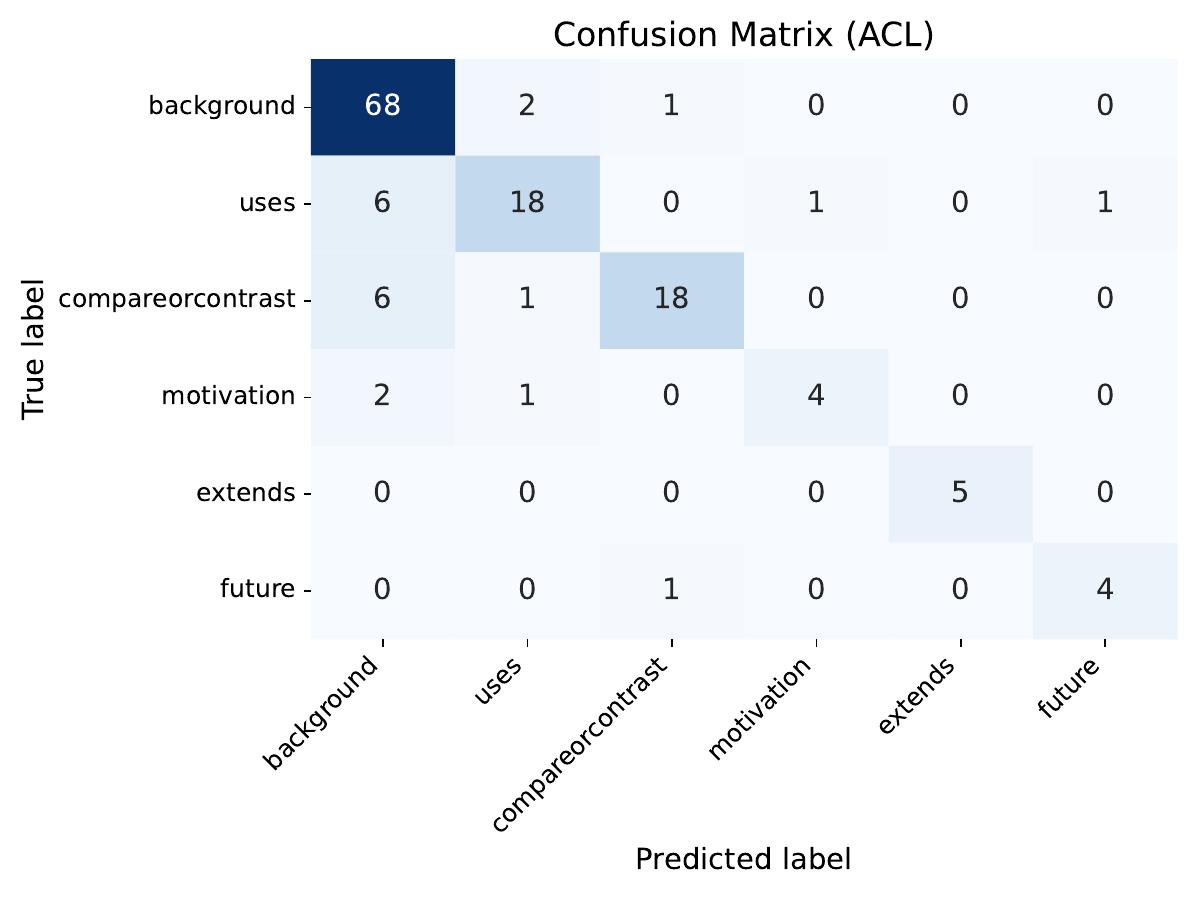} \\
    \vspace{1em}
    \caption{Confusion matrix for SS-cGAN + SciBERT on the ACL dataset.}
    \label{fig:confusion_matrix_acl}
\end{figure}

On the SciCite dataset, the SS-cGAN + SciBERT model achieved a performance of 88.74, which is marginally lower than ImpactCite (88.93), but still competitive. The close performance between these two models indicates that both are highly effective at capturing citation intent in this context. However, it is important to note that direct comparisons cannot be made for the ACL and 3C datasets, as ImpactCite results were not available for those datasets. For the ACL dataset, our semi-supervised SS-cGAN + SciBERT model achieved an F$_1$ score of 81.75\%, significantly outperforming baseline methods such as CitePrompt (68.39\%) and SciBERT Finetune (70.98\%). To further illustrate its strong generalization capability, Figure~\ref{fig:confusion_matrix_acl} shows the confusion matrix of the SS-cGAN + SciBERT predictions, where the vast majority of examples are correctly classified across all citation intent categories.

On the 3C dataset, however, the SS-cGAN + SciBERT model's performance was considerably lower than expected. The model achieved an F1 score of 26.22 (public) and 23.21 (private), significantly trailing behind both the model proposed by \cite{cohan2019structural} (30.3 public / 26.1 private) and the best submission in the 3C Shared Task competition (33.9 public / 27 private). This discrepancy could be attributed to the unique challenges posed by the 3C dataset, which may require more specialized techniques or domain-specific adaptations not fully captured by the SS-cGAN + SciBERT framework. In terms of other baseline models, CitePrompt (86.33 for SciCite) and VarMAE (86.32 for SciCite) both performed reasonably well but were outperformed by the SS-cGAN + SciBERT method in the SciCite dataset, with the latter also showing better results on the ACL dataset. 

We replaced SciBERT with XLNet—the architecture previously adopted in ImpactCite—to closely align our experimental approach with prior work. This transition required substantial modifications to the source code, along with careful adjustments to several model parameters. However, despite these adaptations, the results using XLNet were not as favorable as anticipated. It is worth highlighting that while the reference material clearly outlines the general strategy—fine-tuning XLNet using pre-trained weights and splitting the SciCite dataset—it does not provide explicit details on essential hyperparameters. Key parameters such as learning rate, the exact number of training epochs, batch size, and the chosen optimizer (e.g., Adam) were omitted, making precise replication of ImpactCite's performance challenging.

{A key advantage of our method is its computational efficiency when compared to ImpactCite. While XLNet-large contains approximately 340 million parameters, our SciBERT-based model, augmented with discriminator (50 K parameters) and generator (1M parameters) modules, comprises about 110 million parameters during inference, less than half the size. This significant reduction translates to faster inference times and reduced resource requirements, advantageous for practical deployment scenarios. In conclusion, although our SS-cGAN + SciBERT model demonstrated promising performance and computational efficiency, further refinements are necessary for improved generalization, particularly on more challenging datasets like 3C.}

\subsection{Analysis Model Performance}

The analyses are constrained to training the model using the SciCite dataset, as we initially utilize the Background, Method, and Result classes to perform inferences on the UnarXiv dataset. Subsequently, we will evaluate the impact of citations containing these intents by applying centrality measures.

\begin{figure}[!htb]
    \centering
    \includegraphics[width=0.79\textwidth]{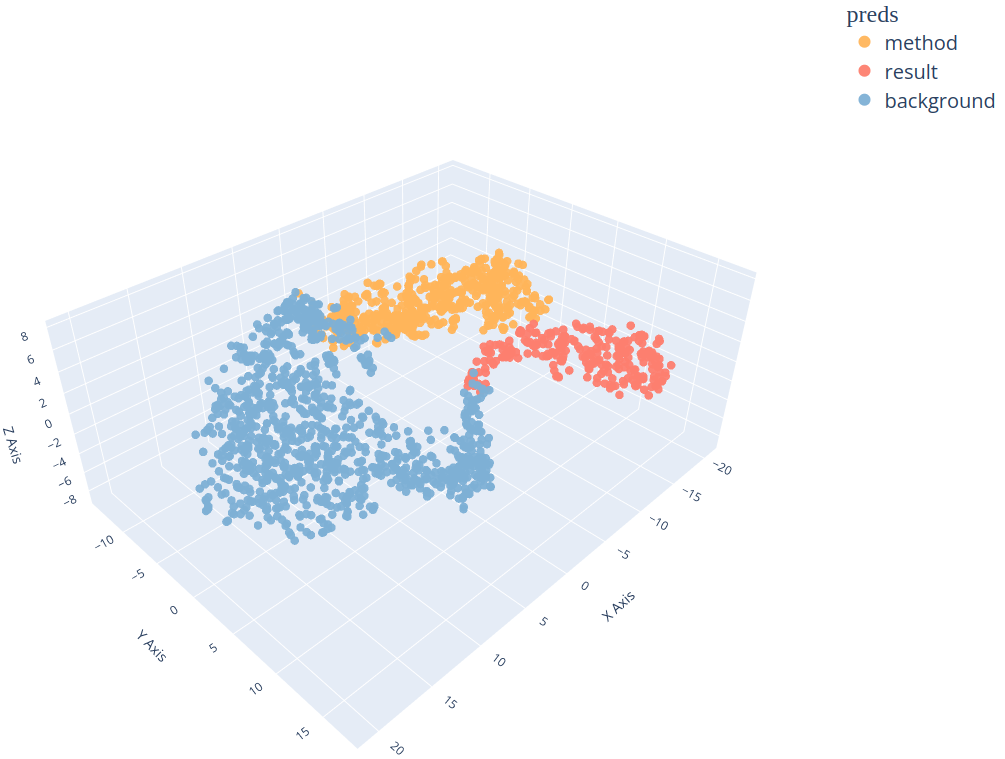}
    \caption{t-SNE visualization of \emph{predicted citation intents} for the SciCite dataset after dimensionality reduction. Each point represents a citation context, with colors indicating different citation intents.}
    \label{fig:predict_representations}
    \vspace{0.5em}
\end{figure}

\begin{figure}[!htb]
    \centering
    \includegraphics[width=0.79\textwidth]{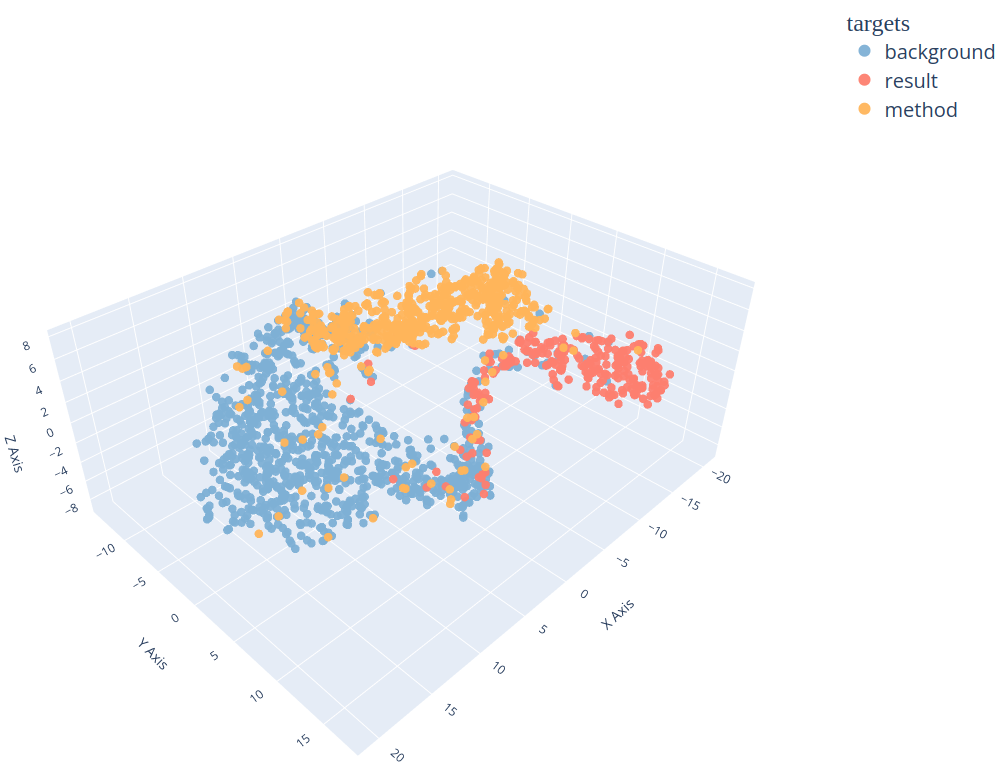}
    \caption{t-SNE visualization of \emph{ground truth citation intents} for the SciCite dataset after dimensionality reduction. Each point represents a citation context, with colors indicating different citation intents.}
    \label{fig:target_representations}
    \vspace{0.5em}
\end{figure}

The t-distributed Stochastic Neighbor Embedding (t-SNE) visualizations in Figures \ref{fig:predict_representations} and \ref{fig:target_representations} depict the SciBERT model embeddings of the SciCite dataset, showing the predicted and target values for the test samples, respectively.
Each point in the image represents a citation context, with each color corresponding to a citation intent. In Figure \ref{fig:predict_representations}, the classes appear relatively well separated, although some overlap, particularly between the Background and Method categories is observed. 
Figure \ref{fig:target_representations} presents the same graph, this time using the target labels. A greater degree of class overlap is observed, with the Background and Method categories being particularly difficult to distinguish.
This overlap suggests that the model struggles to differentiate between these two similar classes.

Despite these challenges, the predicted embeddings exhibit a clearer separation, indicating that the model performs well in distinguishing citation intents at a high level, even though it faces difficulties in making finer distinctions. Further refinement of the model, particularly in differentiating closely related classes, may enhance its performance. 

\begin{figure}[htb]
    \centering
    \includegraphics[width=100mm,scale=0.5]{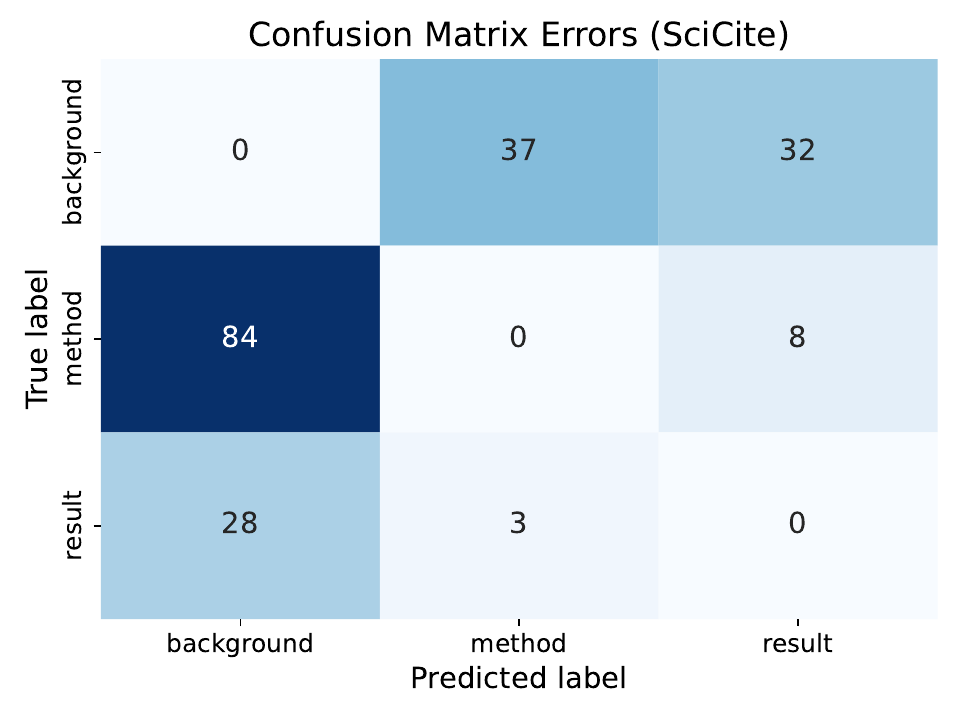} \\
    \vspace{1em}
    \caption{Confusion Matrix displaying classification errors of the SciCite model.}
    \label{fig:confusion_matrix}
\end{figure}

To gain deeper insights into cases where the model exhibited higher error rates, we conducted a more in-depth qualitative analysis.
Out of the 1,861 examples evaluated, 192 were misclassified, accounting for approximately 10\% of the total. Among these errors, the majority occurred when attempting to classify an intention belonging to the Method class, which was incorrectly assigned to the Background class. Approximately 43\% of the total errors are attributed to this confusion between Method and Background, with some additional errors occurring when instances of Background were misclassified as Method. Figure \ref{fig:confusion_matrix} presents the confusion matrix containing only the errors made by the model.

\subsection{Understanding the Impact of Filtering by Citation Intent}

To better understand the influence of different citation intentions on the structure and dynamics of citation networks, we conducted an analysis using complex network measures on the citation network constructed from the unarXiv dataset. Prior to the analysis, two main steps were performed: data preparation and inference execution. 

The unarXiv dataset is provided in JSONL format. Therefore, the first step involved extracting essential information, such as the citation, the article title, the section in which the citation occurred, and other relevant data for our analysis. Next, the data were converted into CSV format to facilitate reading and preparation for inference. For the classification of citation intentions in the unarXiv dataset, we used a model previously trained on the SciCite dataset.

From a total of 149,375 citations, the inference process yielded the following distribution: 85,535 (\(\sim 57\%\)) were classified as \textit{background}, 59,272 (\(\sim 40\%\)) as \textit{method}, and 4,568 (\(\sim 3\%\)) as \textit{result}. This distribution is similar to that observed in the SciCite dataset used for model training, where 58\% of the citations are classified as \textit{background}, 29\% as \textit{method}, and 14\% as \textit{result}.

\begin{figure}[htb]  
    \centering  
    \includegraphics[width=150mm,scale=0.5]{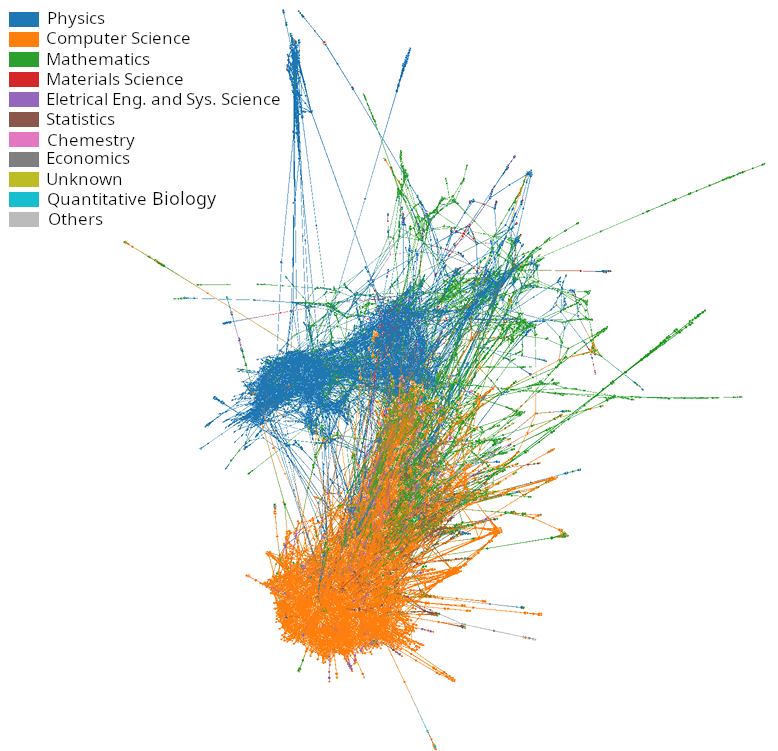}  
    \caption{    \label{fig:citation_network_gwc}  Representation of the largest weakly connected subgraph. Most of the network is dominated by Physics Computer Science and Mathematics papers.} 
\end{figure}

We constructed a citation network comprising 76,640 nodes and 171,403 edges. Our analysis focused on the largest component, extracted from the weakly connected subgraph, which consists of 71,939 nodes and 165,615 edges.
Figure \ref{fig:citation_network_gwc} displays the resulting graph, where a strong interaction between the fields of Computer Science, Physics, and Mathematics can be observed.  

To examine the structural impact of removing citations based on their intent, we performed a filtering experiment on the original citation network. Table \ref{tab:network_filtering_impact} summarizes the effects of filtering each citation intent type individually.
The results show that citations classified as \textit{background} have the greatest impact on network cohesion, significantly reducing the number of nodes (by approximately 51\%) and edges (by approximately 62\%) and dramatically increasing fragmentation into disconnected components (by nearly 567\%). This indicates that \textit{background} citations serve as critical structural links, ensuring the integrative coherence of the scholarly network.

\begin{table}[htb]
    \centering
    \begin{tabular}{l c c c c}
        \hline
        \textbf{Metric}      & \textbf{Full Network} & \textbf{Background Filtered} & \textbf{Method Filtered} & \textbf{Result Filtered} \\ \hline
        Nodes                & 76,640                & 37,644 (–50.9\%)             & 70,877 (–7.5\%)          & 76,115 (–0.7\%)          \\ \hline
        Edges                & 171,403               & 65,739 (–61.6\%)             & 137,850 (–19.6\%)        & 170,070 (–0.8\%)         \\ \hline
        Components           & 340                   & 2,267 (+566.8\%)             & 511 (+50.3\%)            & 347 (+2.1\%)             \\ \hline
    \end{tabular}
    \caption{Impact of filtering citations by intent on network structure.}
    \label{tab:network_filtering_impact}
\end{table}

Filtering \textit{method} citations also notably affected network connectivity, albeit less severely than background citations. The removal resulted in a 7.5\% decrease in nodes, nearly a 20\% reduction in edges, and a 50\% increase in the number of components, underscoring the importance of methodological citations in maintaining connectivity across research areas.

Conversely, citations labeled as \textit{result} had minimal structural impact. Their removal barely changed the number of nodes and edges (less than 1\%) and only slightly increased network fragmentation (by about 2\%), suggesting that result-oriented citations contribute minimally to the structural integrity of the citation network.

These findings highlight the distinct roles that different citation intents play within scholarly networks and emphasize the critical function of \textit{background} and, to a lesser extent, \textit{method} citations in maintaining the structural cohesion and stability of citation networks.

The global effects of filtering intentions observed in the network might be driven by the type of citation intent, as filtering the most frequent citation intents can produce a substantial impact on the overall structure.
However, a different pattern may emerge when the network is analyzed from a local perspective. To illustrate this, we examine the influence of filtering citation intents on network centrality metrics. Specifically, our analysis focuses on the removal of background citations, since these may represent generic or foundational references that do not reflect a direct methodological or conceptual influence. We then investigate how this filtering process affects the ranking of the most central papers in the network.

In our analysis, we examined changes in paper rankings based on centrality measures -- in-degree, closeness, PageRank, and betweenness -- before and after filtering. These changes were visualized using bump charts, highlighting the top 20 papers identified by each centrality metric.

\begin{figure}[htb]
  \centering
  \includegraphics[width=0.9\textwidth]{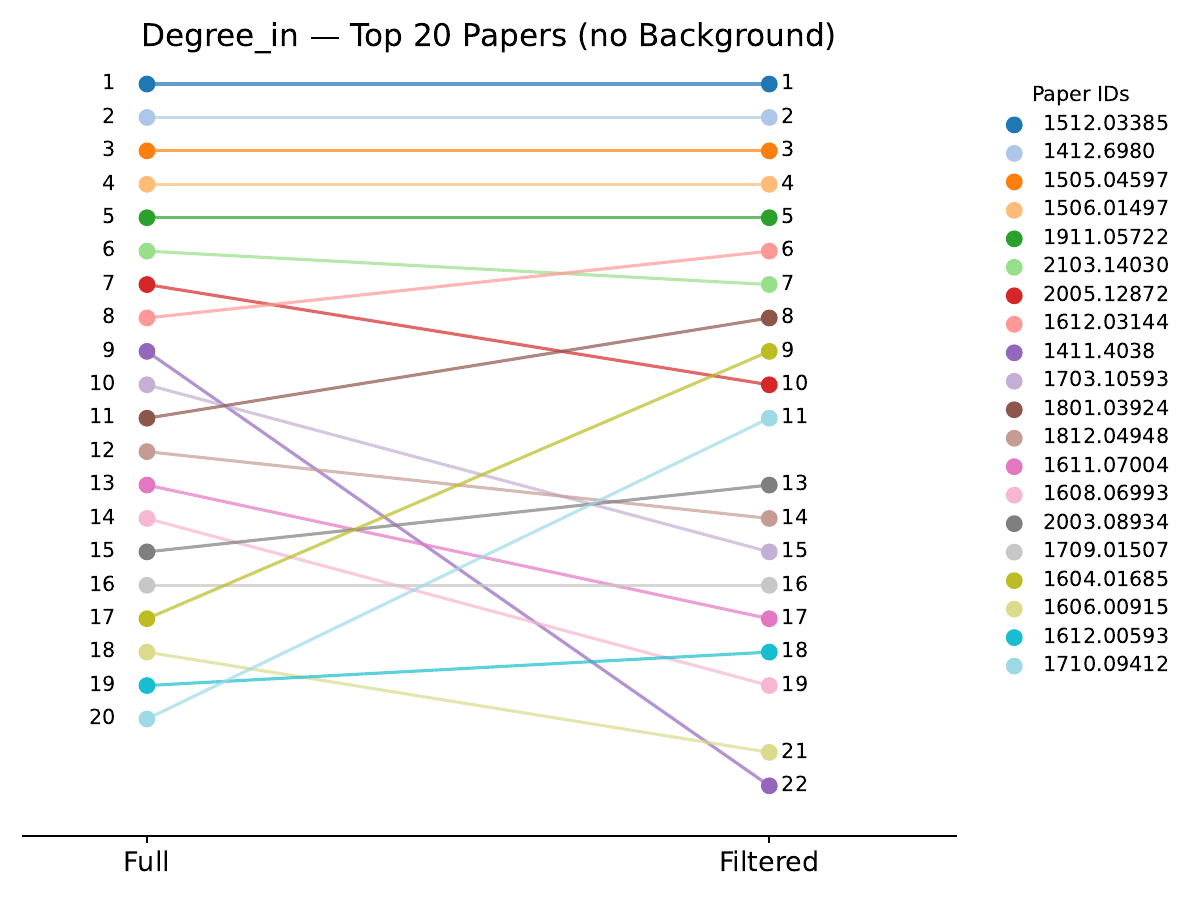}
  \caption{Change in the \emph{in-degree} centrality rankings of the top 20 papers before and after filtering out background citations. Note how paper 1411.4038 drops from rank 9 to 22, while paper 1612.03144 rises from rank 8 to 6.}
  \label{fig:degree_in_no_background}
\end{figure}

Figure \ref{fig:degree_in_no_background} illustrates that, after removing background citations, the in-degree centrality analysis exhibits significant shifts in the rankings of the most cited papers. Notably, paper 1411.4038 
dramatically decreased in rank from 9 to 22, indicating that its previously high centrality relied significantly on background citations. Conversely, paper 1612.03144 
moved up from rank 8 to 6, highlighting an increased importance once the network was filtered.

\begin{figure}[htbp]
  \centering
  \includegraphics[width=0.9\textwidth]{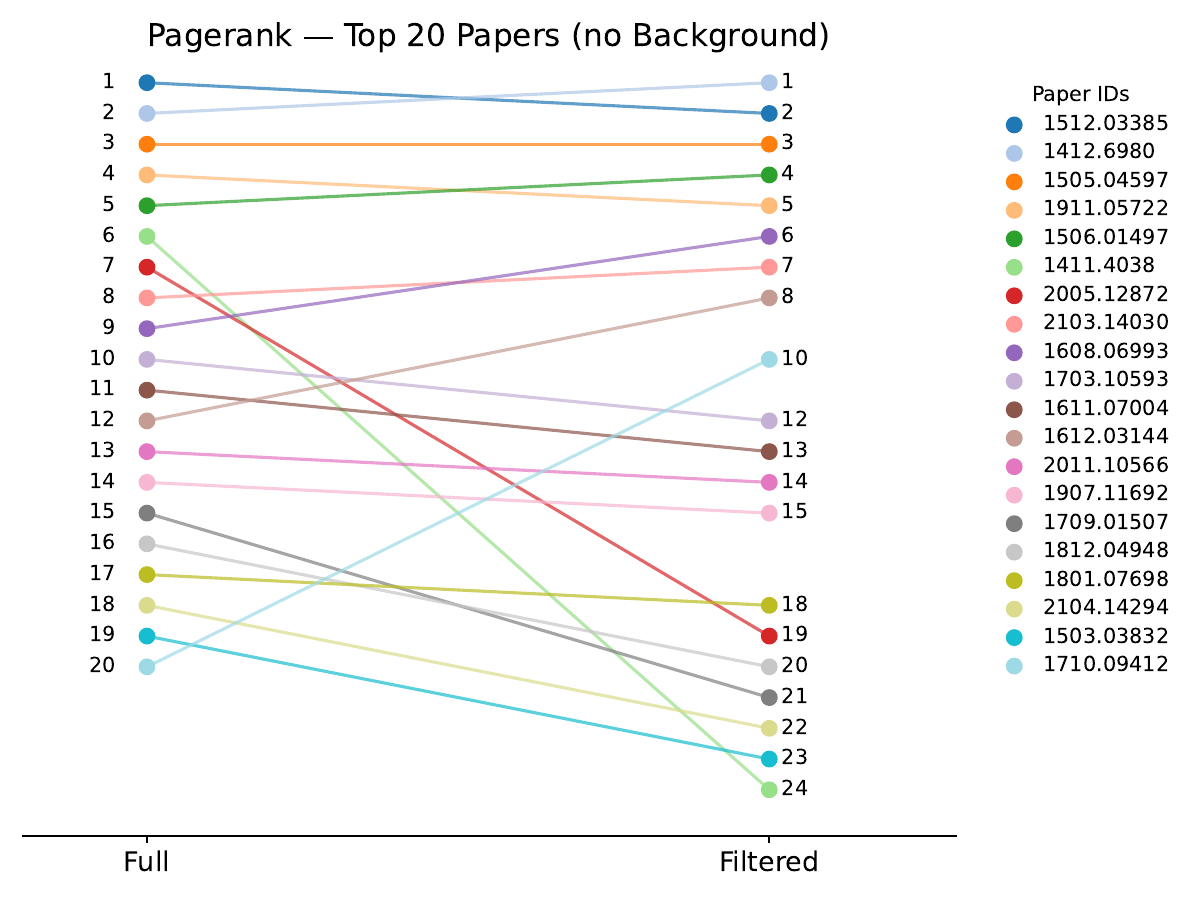}
  \caption{Changes in \emph{PageRank} centrality rankings for the top 20 papers before and after background citation filtering. Paper 1412.6980 rises from rank 2 to 1, indicating increased centrality independent of background citations, while paper 1411.4038 experiences a sharp decline from rank 6 to 24, suggesting strong dependence on background-oriented connections.}
  \label{fig:pagerank_no_background}
\end{figure}

For PageRank centrality (see Figure~\ref{fig:pagerank_no_background}), the rankings of the top papers also showed substantial variation. Paper 1412.6980 rose from rank 2 to become the most central paper after filtering, suggesting that its influence is largely independent of background citations. In contrast, paper 1411.4038 experienced a notable drop, falling from rank 6 to 24, further indicating its dependence on background-oriented connections.

\begin{figure}[htbp]
  \centering
  \includegraphics[width=0.9\textwidth]{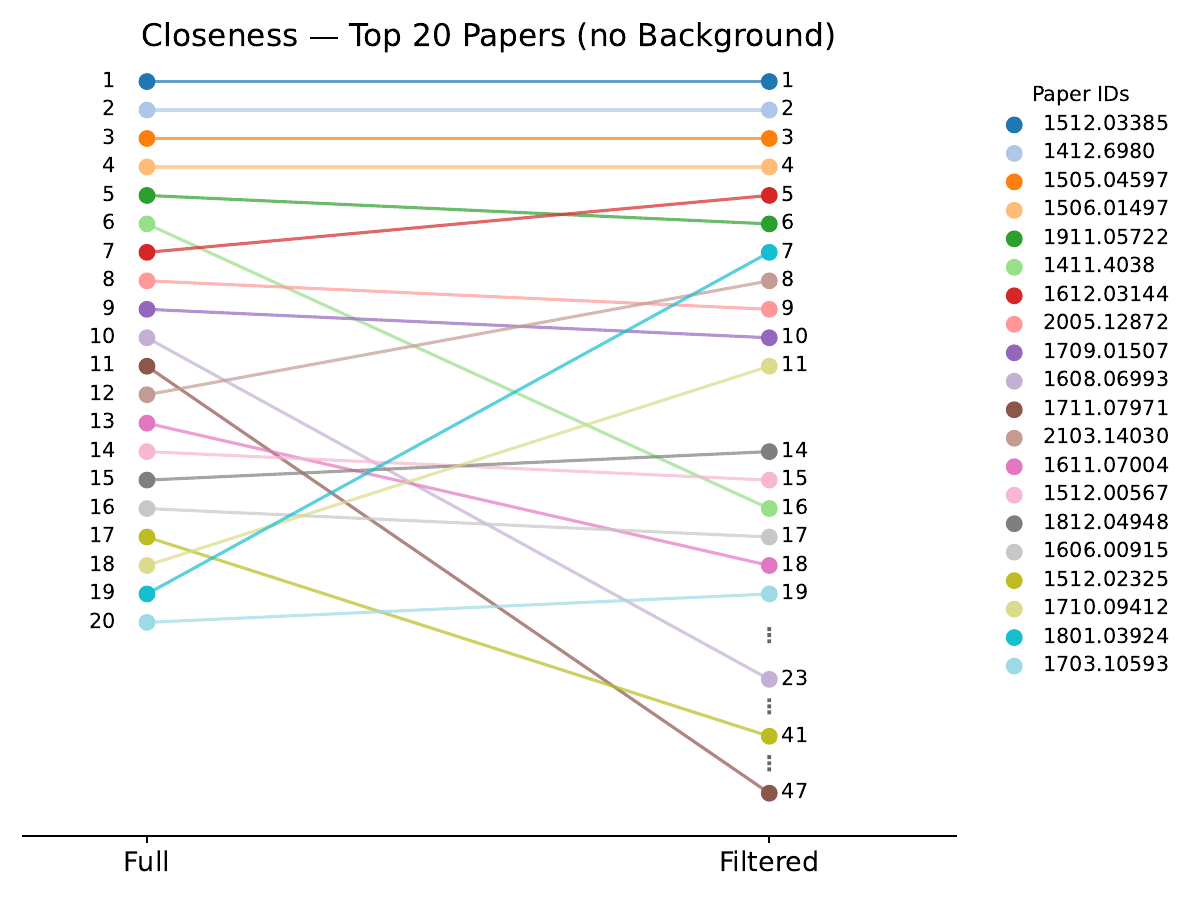}
  \caption{Bump chart comparing \emph{closeness} centrality rankings of the top 20 papers before and after removing background citations. Paper 1711.07971 declines from rank 11 to 47, while paper 1801.03924 improves from rank 19 to 7.}
  \label{fig:closeness_no_background}
\end{figure}

Closeness centrality (see Figure~\ref{fig:closeness_no_background}) also revealed noteworthy shifts. Paper 1711.07971 experienced a sharp decline in rank—from 11 to 47—indicating a strong reliance on background citations to maintain its central position. In contrast, paper 1801.03924 saw a significant improvement, rising from rank 19 to 7, highlighting an increase in its relative centrality after the removal of background citations.

Finally, betweenness centrality displayed the most pronounced changes (result not shown), likely due to its higher sensitivity to structural perturbations in the network. For example, paper 2008.00714, which was initially ranked 12th, experienced a drastic drop, virtually disappearing from the top rankings. In contrast, paper 1801.07698 rose significantly -- from rank 7 to rank 1 -- emerging as the most central paper following the removal of background citations.

The results show, as a proof of principle, that when certain citation intents are filtered out, the relative relevance of papers may change significantly. This shift can influence evaluation metrics used in bibliometric analyses, potentially affecting how authors are assessed and compared. These findings illustrate how a more specific and contextualized analysis of citations can reveal alternative perspectives, highlighting nuanced patterns of scholarly influence that are often overlooked in traditional approaches.

\section{Conclusion }
\label{Conclusion}

Citations have traditionally been used in quantitative research and citation network analysis, often neglecting their rhetorical and functional roles. Common examples of such quantitative approaches include metrics like citation counts, the h-index, disruption index, and impact factor. This paper addresses that limitation by investigating how citation intent -- specifically the distinction between background, methodological, and result-oriented citations -- affects both the structure and interpretation of citation networks.

{As our first contribution, we adopt a semi-supervised model, named \emph{cGAN-SciBERT}, for the task of citation intent classification. By leveraging SciBERT’s capability to represent contextual information and the discriminative power of GANs, the model achieves competitive performance on benchmark datasets, particularly \textsc{SciCite} and \textsc{ACL-ARC}. The approach remains effective even when trained on limited labeled data. Our semi-supervised model attains an F$_1$ score of 88.74\,\% on the \textsc{SciCite} benchmark -- only 0.22 pp below the XLNet-large-based \textsc{ImpactCite}’s 88.93\,\% -- while using fewer than half the parameters ($\approx110$\,M vs.\ 340\,M). This demonstrates that the GAN framework can closely match state-of-the-art classification performance with substantially reduced inference cost, making our approach both accurate and highly efficient.}

Our second contribution involved the use of citation intent filtering to analyze the structure of citation networks. We constructed citation networks, selectively removed specific citation intents, and examined the impact on various centrality metrics, including in-degree, PageRank, closeness, and betweenness. The results demonstrated that filtering by citation intent can lead to substantial changes in centrality-based rankings. This suggests that traditional bibliometric indicators, which do not account for citation intent, may obscure meaningful distinctions in scholarly influence. Papers that ranked highly in the unfiltered analysis experienced significant drops in ranking when intent-based filtering was applied. Conversely, papers that initially held lower positions in the traditional analysis improved their rankings under intent-based filtering. Significant changes in ranking were observed in simpler metrics like in-degree centrality, and to an even greater extent in betweenness centrality.

Future work could explore how citation intentions can refine bibliometric indicators such as the disruption index. By distinguishing between different intents one could construct intent-aware variants of the disruption metric that more accurately reflect whether a paper introduces genuinely novel ideas or builds upon existing frameworks. For example, a high proportion of method or background citations may indicate continuity rather than disruption.
Also, future work could investigate whether certain linguistic cues in citation contexts signal that a paper is perceived as disruptive when cited.
Additionally, intent-based analysis may support bias and ethics monitoring by identifying strategic citation practices, such as excessive self-citation in background sections or citation inflation through non-substantive references. This would enable a more precise and transparent understanding of scholarly influence and research behavior.



%

\newpage

\section*{Supplementary Information}

This supplementary material provides details on the construction of the citation networks and the centrality metrics used to assess paper relevance (see Section \ref{citneta}). Section \ref{sec:training-configuration} outlines the parameters used for training the GANs.

\subsection{Citation Networks and Centrality Metrics} \label{citneta}

Citation networks are a valuable tool for analyzing the flow of academic knowledge through directed links. They are frequently used to investigate patterns of scholarly influence, the structure of scientific knowledge, and the dynamics of research dissemination \cite{sciencedynamics2012andrea}.
The structure of citation networks often follows 
patterns found in many real-world networks, such as 
small-world properties and scale-free distributions \cite{barabasi1999emergence}.


A citation network is a directed graph \( G = (V, E) \), 
where each node \( v_i \in V \) represents an academic paper, 
and an edge \( (v_i, v_j) \in E \) indicates that paper \( v_i \) 
cites paper \( v_j \). This structure captures the flow of information 
within the scientific community, allowing for the analysis of 
relationships among scholarly works. 
Citation networks are typically sparse, with most papers having 
few citations, while a small number of highly influential ``hub'' 
papers accumulate numerous citations, shaping the network’s hierarchical, 
scale-free topology.

Centrality measures are fundamental for identifying influential papers and authors within citation networks. These metrics, grounded in social network analysis, evaluate the structural importance of nodes by analyzing how information flows between publications~\cite{betweenness2007loet}. In citation networks, centrality provides insights into which articles exert the greatest impact, which researchers are most influential, and how knowledge disseminates. Traditionally, these measures aim to identify nodes that occupy critical structural positions, although assessing the distinctiveness of such centrality can be challenging \cite{freeman1978}.

Several centrality measures have been developed to assess the relevance of nodes within a network. Betweenness centrality highlights nodes that frequently appear on the shortest paths between others, revealing their role as critical intermediaries in information flow~\cite{freeman1978, centrality2007leydesdorff}. Another example is the closeness centrality, which identifies nodes that are, on average, closer to all others in the network, reflecting their efficiency in disseminating information~\cite{freeman1978}. These metrics offer valuable insights into the dynamics of scientific influence, capturing both the accumulation of citations and the structural relationships across journals.

The centrality metrics employed in this study are described below.

\begin{enumerate}

\item \emph{Degree}: this is the most basic centrality measure and refers to the number of links (or edges) a node has in the network. In citation networks, the degree centrality of a paper can be split into two types: \emph{in-degree} $(k_{\mathrm{in}})$  and \emph{out-degree} $(k_{\mathrm{out}})$.
To assess the importance of a paper, the most commonly used metric is in-degree. The in-degree  corresponds to the number of citations a paper receives. Papers with a high in-degree are often considered influential because they have been cited by many other works.

Mathematically, for a directed graph with adjacency matrix \( A \), where \( a_{ij} = 1 \) if paper \( i \) cites paper \( j \) (and 0 otherwise), the in-degree and out-degree are given by the following sums:

\begin{equation}
    \centering
    k_{\mathrm{in}}(i) = \sum_{j} a_{ji}, \quad k_{\mathrm{out}}(i) = \sum_{j} a_{ij}.
\end{equation}

\item \emph{Betweenness}: this metric  quantifies the role of a node in facilitating the flow of information within a network by measuring how often it appears on the shortest paths between other nodes. In the context of citation networks, where papers \( V \) are connected through citation links \( E \), betweenness centrality identifies papers that act as critical intermediaries, often bridging distinct research areas. Such papers serve as ``glue'' within the network, connecting otherwise disparate studies and enabling the dissemination of ideas across disciplines.

The betweenness centrality of a paper \( v \) is defined as:
\begin{equation}
    \centering
    g(v) = \sum_{s \neq t \neq v} \frac{\sigma_{st}(v)}{\sigma_{st}}
\end{equation}
where \( \sigma_{st} \) represents the total number of shortest paths between papers \( s \) and \( t \), and \( \sigma_{st}(v) \) is the number of these paths that pass through paper \( v \). A paper with high betweenness centrality frequently appears on these shortest paths, highlighting its role in transferring knowledge between different parts of the network.

Papers with high betweenness centrality are pivotal in the structure of citation networks. They not only facilitate communication between specialized fields but also play a key role in the integration of diverse scientific ideas, thus enhancing the overall connectivity and coherence of the academic landscape.

\item \emph{Closeness}: This centrality measure captures how efficiently a node can disseminate information throughout the network. In citation networks, where papers are connected through citation links, those with high closeness centrality tend to be more easily discoverable by an agent navigating the network.
Formally, the closeness centrality of a paper \( v \) is defined as:
\begin{equation}
    \centering
    g(v) = \sum_{u \in N, u \neq v} \frac{\gamma(u, v)}{N}
\end{equation}
where \( \gamma(u, v) \) represents the minimum number of papers and citations needed to traverse the network from paper \( u \) to paper \( v \), and \( N \) is the total number of papers in the network. Closeness centrality is inversely related to the average shortest path distance.

\item \emph{PageRank}: Originally introduced by Brin and Page (1998) \cite{brin1998anatomy} to rank web pages based on their importance and authority, PageRank has found direct applicability in citation networks due to the similarity between web hyperlinks and academic citations. Unlike simple in‐degree centrality, which treats all inbound citations equally, PageRank assigns importance to citations based on the prestige and citation behavior of the citing papers. Specifically, the PageRank algorithm calculates the importance of a paper \(A\) recursively by considering the weighted influence of papers \(T_i\) citing it, normalized by their number of outbound citations \(C(T_i)\):
\begin{equation}
  \mathrm{PR}(A) \;=\;(1 - d)\;+\;d \sum_{T_i \rightarrow A} \frac{\mathrm{PR}(T_i)}{C(T_i)}.
\end{equation}
Here, \(\mathrm{PR}(T_i)\) denotes the PageRank of the citing paper \(T_i\), \(C(T_i)\) is the number of outbound citations from \(T_i\), and \(d\) is a damping factor typically set between 0 and 1. The damping factor represents the probability of continuing to follow citation chains versus randomly jumping to another paper.

Thus, PageRank centrality provides a more nuanced and robust measure of importance in citation networks, effectively incorporating both the quantity and quality (prestige) of citations. Papers receiving citations from highly influential sources will achieve higher PageRank scores, reflecting their increased visibility and authoritative status within the scientific community.
\end{enumerate}

\subsection{Hyperparameter Configuration for GAN Training}
\label{sec:training-configuration}
\begin{enumerate}

\item \emph{SciCite with cGAN-SciBERT}: The semi-supervised GAN‐BERT model built on SciBERT was trained on the SciCite benchmark. We fixed most architectural choices to match SciBERT’s pretrained configuration, and tuned only the GAN-specific and optimization hyperparameters. The best validation performance was observed at epoch 16. Table \ref{tab:hyperparams-scicite} lists the hyperparameters employed for training cGAN-SciBERT on the SciCite dataset.

\begin{table}[ht]
  \centering
  \begin{tabular}{l l}
    \hline
    \textbf{Parameter}                      & \textbf{Value} \\
    \hline
    Maximum sequence length ($L_{\max}$)    & 160           \\
    Batch size                              & 32            \\
    Generator hidden layers ($G$)           & 1             \\
    Discriminator hidden layers ($D$)       & 1             \\
    Noise vector dimensionality ($z$)       & 768           \\
    Dropout rate (output)                   & 0.20          \\
    Learning rate (discriminator)           & $2\times10^{-7}$ \\
    Learning rate (generator)               & $2\times10^{-7}$ \\
    Adam $\epsilon$                         & $2\times10^{-7}$ \\
    Number of epochs                        & 20            \\
    Warm-up proportion                      & 0.10          \\
    Pretrained model name                   & \texttt{allenai/scibert\_scivocab\_uncased} \\
    Hidden size (SciBERT)                   & 768           \\
    \hline
  \end{tabular}
  \caption{Hyperparameters for training cGAN-SciBERT on the SciCite dataset.}
  \label{tab:hyperparams-scicite}
\end{table}

\item \emph{ACL and 3C with cGAN-SciBERT}: For the smaller ACL and 3C corpora, we reduced sequence length and adjusted the GAN architecture and learning rates to account for dataset size. The optimal stopping epoch was chosen by monitoring the validation F$_1$. Table \ref{tab:hyperparams-acl3c} lists the hyperparameters employed for training cGAN-SciBERT on ACL and 3C datasets.

\begin{table}[ht]
  \centering
  \begin{tabular}{l l}
    \hline
    \textbf{Parameter}                      & \textbf{Value} \\
    \hline
    Maximum sequence length ($L_{\max}$)    & 64            \\
    Batch size                              & 16            \\
    Generator hidden layers ($G$)           & 2             \\
    Discriminator hidden layers ($D$)       & 1             \\
    Noise vector dimensionality ($z$)       & 100           \\
    Dropout rate (output)                   & 0.10          \\
    Learning rate (discriminator)           & $5\times10^{-5}$ \\
    Learning rate (generator)               & $5\times10^{-4}$ \\
    Adam $\epsilon$                         & $2\times10^{-7}$ \\
    Number of epochs                        & 30            \\
    Warm-up proportion                      & 0.10          \\
    Pretrained model name                   & \texttt{allenai/scibert\_scivocab\_uncased} \\
    Hidden size (SciBERT)                   & 768           \\
    \hline
  \end{tabular}
  \caption{Hyperparameters for training cGAN-SciBERT on ACL and 3C datasets.}
  \label{tab:hyperparams-acl3c}
\end{table}

\end{enumerate}

\newpage

\bibliographystyle{ieeetr}
\bibliographystyle{abbrv}

\begin{thebibliography}{10}

\bibitem{svm_naive_bayes_2010}
S.~Agarwal, L.~Choubey, and H.~Yu.
\newblock Automatically classifying the role of citations in biomedical articles.
\newblock {\em AMIA ... Annual Symposium proceedings / AMIA Symposium. AMIA Symposium}, 2010:11--5, 11 2010.

\bibitem{auti_etal_2022}
T.~Auti, R.~Sarkar, B.~Stearns, A.~K. Ojha, A.~Paul, M.~Comerford, J.~Megaro, J.~Mariano, V.~Herard, and J.~P. McCrae.
\newblock Towards classification of legal pharmaceutical text using {GAN}-{BERT}.
\newblock In M.~Wan and C.-R. Huang, editors, {\em Proceedings of the First Computing Social Responsibility Workshop within the 13th Language Resources and Evaluation Conference}, pages 52--57, Marseille, France, June 2022. European Language Resources Association.

\bibitem{bagrow2024working}
J.~Bagrow and Y.-Y. Ahn.
\newblock {\em Working with Network Data}.
\newblock Cambridge University Press, 2024.

\bibitem{barabasi1999emergence}
A.-L. Barabási and R.~Albert.
\newblock Emergence of scaling in random networks.
\newblock {\em Science}, 286(5439):509--512, 1999.

\bibitem{beltagy_etal_2019_scibert}
I.~Beltagy, K.~Lo, and A.~Cohan.
\newblock {S}ci{BERT}: A pretrained language model for scientific text.
\newblock In K.~Inui, J.~Jiang, V.~Ng, and X.~Wan, editors, {\em Proceedings of the 2019 Conference on Empirical Methods in Natural Language Processing and the 9th International Joint Conference on Natural Language Processing (EMNLP-IJCNLP)}, pages 3615--3620, Hong Kong, China, Nov. 2019. Association for Computational Linguistics.

\bibitem{bird_etal_2008_acl}
S.~Bird, R.~Dale, B.~Dorr, B.~Gibson, M.~Joseph, M.-Y. Kan, D.~Lee, B.~Powley, D.~Radev, and Y.~F. Tan.
\newblock The {ACL} {A}nthology reference corpus: A reference dataset for bibliographic research in computational linguistics.
\newblock In N.~Calzolari, K.~Choukri, B.~Maegaard, J.~Mariani, J.~Odijk, S.~Piperidis, and D.~Tapias, editors, {\em Proceedings of the Sixth International Conference on Language Resources and Evaluation ({LREC}'08)}, Marrakech, Morocco, May 2008. European Language Resources Association (ELRA).

\bibitem{brin1998anatomy}
S.~Brin and L.~Page.
\newblock The anatomy of a large-scale hypertextual web search engine.
\newblock {\em Computer Networks}, 30:107--117, 1998.

\bibitem{brito2023analyzing}
A.~C. Brito, F.~N. Silva, and D.~R. Amancio.
\newblock Analyzing the influence of prolific collaborations on authors productivity and visibility.
\newblock {\em Scientometrics}, 128(4):2471--2487, 2023.

\bibitem{brito2020complex}
A.~C.~M. Brito, F.~N. Silva, and D.~R. Amancio.
\newblock A complex network approach to political analysis: Application to the brazilian chamber of deputies.
\newblock {\em Plos one}, 15(3):e0229928, 2020.

\bibitem{gan_cite_2025}
K.~Chatrinan, T.~Noraset, and S.~Tuarob.
\newblock Gan-cite: leveraging semi-supervised generative adversarial networks for citation function classification with limited data.
\newblock {\em Scientometrics}, 130(2):679--703, 2025.

\bibitem{cohan2019structural}
A.~Cohan, W.~Ammar, M.~van Zuylen, and F.~Cady.
\newblock Structural scaffolds for citation intent classification in scientific publications.
\newblock In J.~Burstein, C.~Doran, and T.~Solorio, editors, {\em Proceedings of the 2019 Conference of the North {A}merican Chapter of the Association for Computational Linguistics: Human Language Technologies, Volume 1 (Long and Short Papers)}, pages 3586--3596, Minneapolis, Minnesota, June 2019. Association for Computational Linguistics.

\bibitem{cohan2015matching}
A.~Cohan, L.~Soldaini, and N.~Goharian.
\newblock Matching citation text and cited spans in biomedical literature: a search-oriented approach.
\newblock In R.~Mihalcea, J.~Chai, and A.~Sarkar, editors, {\em Proceedings of the 2015 Conference of the North {A}merican Chapter of the Association for Computational Linguistics: Human Language Technologies}, pages 1042--1048, Denver, Colorado, May{--}June 2015. Association for Computational Linguistics.

\bibitem{correa2019word}
E.~A. Corr{\^e}a~Jr and D.~R. Amancio.
\newblock Word sense induction using word embeddings and community detection in complex networks.
\newblock {\em Physica A: Statistical Mechanics and its Applications}, 523:180--190, 2019.

\bibitem{correa2017patterns}
E.~A. Corr{\^e}a~Jr, F.~N. Silva, L.~d.~F. Costa, and D.~R. Amancio.
\newblock Patterns of authors contribution in scientific manuscripts.
\newblock {\em Journal of Informetrics}, 11(2):498--510, 2017.

\bibitem{croce-etal-2020-gan}
D.~Croce, G.~Castellucci, and R.~Basili.
\newblock {GAN}-{BERT}: Generative adversarial learning for robust text classification with a bunch of labeled examples.
\newblock In D.~Jurafsky, J.~Chai, N.~Schluter, and J.~Tetreault, editors, {\em Proceedings of the 58th Annual Meeting of the Association for Computational Linguistics}, pages 2114--2119, Online, July 2020. Association for Computational Linguistics.

\bibitem{devlin2019bert}
J.~Devlin, M.-W. Chang, K.~Lee, and K.~Toutanova.
\newblock {BERT}: Pre-training of deep bidirectional transformers for language understanding.
\newblock In J.~Burstein, C.~Doran, and T.~Solorio, editors, {\em Proceedings of the 2019 Conference of the North {A}merican Chapter of the Association for Computational Linguistics: Human Language Technologies, Volume 1 (Long and Short Papers)}, pages 4171--4186, Minneapolis, Minnesota, June 2019. Association for Computational Linguistics.

\bibitem{dong-schafer-2011-ensemble}
C.~Dong and U.~Sch{\"a}fer.
\newblock Ensemble-style self-training on citation classification.
\newblock In H.~Wang and D.~Yarowsky, editors, {\em Proceedings of 5th International Joint Conference on Natural Language Processing}, pages 623--631, Chiang Mai, Thailand, Nov. 2011. Asian Federation of Natural Language Processing.

\bibitem{freeman1978}
L.~C. Freeman.
\newblock Centrality in social networks conceptual clarification.
\newblock {\em Social Networks}, 1(3):215--239, 1978.

\bibitem{garfield1965can}
E.~Garfield et~al.
\newblock Can citation indexing be automated.
\newblock In {\em Statistical association methods for mechanized documentation, symposium proceedings}, volume 269, pages 189--192. Citeseer, 1965.

\bibitem{garzone1997automated}
M.~Garzone.
\newblock {\em Automated Classification of Citations Using Linguistic Semantic Grammars [microform]}.
\newblock Canadian theses on microfiche. Thesis (M.Sc.)--University of Western Ontario, 1997.

\bibitem{based_rules_2000}
M.~Garzone and R.~E. Mercer.
\newblock Towards an automated citation classifier.
\newblock In H.~J. Hamilton, editor, {\em Advances in Artificial Intelligence}, pages 337--346, Berlin, Heidelberg, 2000. Springer Berlin Heidelberg.

\bibitem{ghosh_2020}
S.~Ghosh and C.~Shah.
\newblock Identifying citation sentiment and its influence while indexing scientific papers.
\newblock 01 2020.

\bibitem{gupta2024sentiment}
S.~Gupta and A.~Kumar.
\newblock Sentiment-aware enhancements of pagerank-based citation metric, impact factor, and h-index for ranking the authors of scholarly articles, 2024.

\bibitem{hu2022varmae}
D.~Hu, X.~Hou, X.~Du, M.~Zhou, L.~Jiang, Y.~Mo, and X.~Shi.
\newblock Varmae: Pre-training of variational masked autoencoder for domain-adaptive language understanding, 2022.

\bibitem{jebari2021use}
C.~Jebari, E.~Herrera-Viedma, and M.~J. Cobo.
\newblock The use of citation context to detect the evolution of research topics: a large-scale analysis.
\newblock {\em Scientometrics}, 126(4):2971--2989, 2021.

\bibitem{jurgens2018}
D.~Jurgens, S.~Kumar, R.~Hoover, D.~McFarland, and D.~Jurafsky.
\newblock Measuring the evolution of a scientific field through citation frames.
\newblock {\em Transactions of the Association for Computational Linguistics}, 6:391--406, 2018.

\bibitem{tacl_a_2018_acl}
D.~Jurgens, S.~Kumar, R.~Hoover, D.~McFarland, and D.~Jurafsky.
\newblock {Measuring the Evolution of a Scientific Field through Citation Frames}.
\newblock {\em Transactions of the Association for Computational Linguistics}, 6:391--406, 07 2018.

\bibitem{lahiri2023citeprompt}
A.~Lahiri, D.~K. Sanyal, and I.~Mukherjee.
\newblock Citeprompt: Using prompts to identify citation intent in scientific papers, 2023.

\bibitem{betweenness2007loet}
L.~Leydesdorff.
\newblock Betweenness centrality as an indicator of the interdisciplinarity of scientific journals.
\newblock {\em Journal of the American Society for Information Science and Technology}, 58(9):1303--1319, 2007.

\bibitem{centrality2007leydesdorff}
L.~Leydesdorff.
\newblock "betweenness centrality" as an indicator of the "interdisciplinarity" of scientific journals.
\newblock {\em Journal of the American Society for Information Science and Technology}, 58, 07 2007.

\bibitem{liho2008use}
Z.~Li and Y.-S. Ho.
\newblock {Use of citation per publication as an indicator to evaluate contingent valuation research}.
\newblock {\em Scientometrics}, 75(1):97--110, April 2008.

\bibitem{mercier2020impactcite}
D.~Mercier, S.~T.~R. Rizvi, V.~Rajashekar, A.~Dengel, and S.~Ahmed.
\newblock Impactcite: An xlnet-based method for citation impact analysis.
\newblock {\em CoRR}, abs/2005.06611, 2020.

\bibitem{michael1975resultson}
M.~J. Moravcsik and P.~Murugesan.
\newblock Some results on the function and quality of citations.
\newblock {\em Social Studies of Science}, 5(1):86--92, 1975.

\bibitem{moravcsik_1975}
M.~J. Moravcsik and P.~Murugesan.
\newblock Some results on the function and quality of citations.
\newblock {\em Social Studies of Science}, 5(1):86--92, 1975.

\bibitem{Note1}
https://www.kaggle.com/c/3c-shared-task-purpose-v2.

\bibitem{Note2}
https://github.com/IllDepence/unarXive.

\bibitem{Note3}
https://github.com/filipinascimento/helios-web.

\bibitem{knoth_act}
D.~Pride and P.~Knoth.
\newblock An authoritative approach to citation classification.
\newblock In {\em Proceedings of the ACM/IEEE Joint Conference on Digital Libraries in 2020}, JCDL '20, page 337–340, New York, NY, USA, 2020. Association for Computing Machinery.

\bibitem{ritchie2008}
A.~Ritchie, S.~Robertson, and S.~Teufel.
\newblock Comparing citation contexts for information retrieval.
\newblock In {\em Proceedings of the 17th ACM Conference on Information and Knowledge Management}, CIKM '08, page 213–222, New York, NY, USA, 2008. Association for Computing Machinery.

\bibitem{saier-2020-unarXive}
T.~Saier and M.~F{\"{a}}rber.
\newblock {unarXive: A Large Scholarly Data Set with Publications’ Full-Text, Annotated In-Text Citations, and Links to Metadata}.
\newblock {\em Scientometrics}, 125(3):3085--3108, Dec. 2020.

\bibitem{sciencedynamics2012andrea}
A.~Scharnhorst, K.~Borner, and P.~Van~den Besselaar.
\newblock {\em Models of Science Dynamics: Encounters Between Complexity Theory and Information Sciences}.
\newblock 01 2012.

\bibitem{silva2016citationnetwork}
F.~N. Silva, D.~R. Amancio, M.~Bardosova, L.~da~F.~Costa, and O.~N. Oliveira.
\newblock Using network science and text analytics to produce surveys in a scientific topic.
\newblock {\em Journal of Informetrics}, 10(2):487--502, 2016.

\bibitem{silva_etal_2023}
K.~Silva, B.~Can, F.~Blain, R.~Sarwar, L.~Ugolini, and R.~Mitkov.
\newblock Authorship attribution of late 19th century novels using {GAN}-{BERT}.
\newblock In V.~Padmakumar, G.~Vallejo, and Y.~Fu, editors, {\em Proceedings of the 61st Annual Meeting of the Association for Computational Linguistics (Volume 4: Student Research Workshop)}, pages 310--320, Toronto, Canada, July 2023. Association for Computational Linguistics.

\bibitem{small2018}
H.~Small.
\newblock Characterizing highly cited method and non-method papers using citation contexts: The role of uncertainty.
\newblock {\em Journal of Informetrics}, 12(2):461--480, 2018.

\bibitem{swales1986citation}
J.~Swales.
\newblock Citation analysis and discourse analysis.
\newblock {\em Applied Linguistics - APPL LINGUIST}, 7:39--56, 03 1986.

\bibitem{teufel2006annotation}
S.~Teufel, A.~Siddharthan, and D.~Tidhar.
\newblock An annotation scheme for citation function.
\newblock In J.~Alexandersson and A.~Knott, editors, {\em Proceedings of the 7th {SIG}dial Workshop on Discourse and Dialogue}, pages 80--87, Sydney, Australia, July 2006. Association for Computational Linguistics.

\bibitem{teufel2006automatic}
S.~Teufel, A.~Siddharthan, and D.~Tidhar.
\newblock Automatic classification of citation function.
\newblock In D.~Jurafsky and E.~Gaussier, editors, {\em Proceedings of the 2006 Conference on Empirical Methods in Natural Language Processing}, pages 103--110, Sydney, Australia, July 2006. Association for Computational Linguistics.

\bibitem{weinstock_1971}
G.~THOMPSON and Y.~YIYUN.
\newblock Evaluation in the reporting verbs used in academic papers.
\newblock {\em Applied Linguistics}, 12(4):365--382, 12 1991.

\bibitem{valenzuela2015identifying}
M.~A. Valenzuela-Escarcega, V.~A. Ha, and O.~Etzioni.
\newblock Identifying meaningful citations.
\newblock In {\em AAAI Workshop: Scholarly Big Data}, 2015.

\bibitem{white2004citation}
H.~D. White.
\newblock Citation analysis and discourse analysis revisited.
\newblock {\em Applied Linguistics}, 25(1):89--116, 03 2004.

\bibitem{centrality2009yan}
E.~Yan and Y.~Ding.
\newblock Applying centrality measures to impact analysis: A coauthorship network analysis.
\newblock {\em Journal of the American Society for Information Science and Technology}, 60(10):2107--2118, 2009.

\end{thebibliography}

\end{document}